
\input amsppt.sty
\magnification=\magstep1

%
%
%
%
%
%
%

\define\Real{{\Bbb R}}
\define\Comp{{\Bbb C}}

\define\Hess{ \ \hbox{Hess} \ }

\def\qed{\hbox{${\vcenter{\vbox{
    \hrule height 0.4pt\hbox{\vrule width 0.4pt height 6pt
    \kern5pt\vrule width 0.4pt}\hrule height 0.4pt}}}$}}

\def\and{\  \hbox{ and } \ }

\def\0{\bold{0}}




\title Critical Points and Singular Vectors
\endtitle
\topmatter
\title Critical Points of the Product of Powers of Linear
Functions and Families of Bases of Singular Vectors
\endtitle
\author A.Varchenko$^*$ \endauthor
\date September, 1993\enddate
\thanks The author was supported by NSF Grant DMS- -9203929.
\newline $^*$ Department of Mathematics, University of North Carolina
, Chapel Hill, NC 27599, USA
\newline {\it E-mail address:} varchenko\@math.unc.edu
\endthanks
\abstract
The quasiclassical asymptotics of the Knizhnik-Zamolodchikov equation
with values in the tensor product of $sl_2$ representations are
considered. The first term of asymptotics is an eigenvector of a system
of commuting operators. We show that the norm of this vector with
respect to the Shapovalov form is equal to the determinant of the matrix
of second derivatives of a suitable function. This formula is an analog
of the Gaudin and Korepin formulae for the norm of the Bethe vectors. We
show that the eigenvectors form   a basis under certain conditions.
\endabstract
\endtopmatter

\document
\subhead Introduction \endsubhead
\par Consider the Lie algebra  $sl_2$ with standard generators
$e, f, h$ such that $[e,f]=h, [h,e]=2e, [h,f]=-2f$. Let $L^n$ be the
$(n+1)-$dimensional irreducible $sl_2$ module.
The module is generated by its singular element $v_n$ such
that $ev_n=0$ and $hv_n=nv_n$. The elements $v_n, fv_n,...
,f^nv_n$ form a basis of $L^n$.
 The Shapovalov form on $L^n$ is the bilinear form $B^n$ such that
$$
B^n(f^kv_n,f^kv_n)=k!n!/(n-k)!,\ \ \ \ \
B^n(f^kv_n,f^lv_n)=0   \hbox{\ for\ } k \neq l.
$$
  The tensor product of irreducible representations is direct sum of
irreducible representations:
  $L^n \otimes L^m=L^{m-n} \oplus L^{m-n+2} \oplus ... \oplus L^{m+n}$ for
$m \geq n$,
   and a singular vector of $L^{m+n-2k}$ has the form
$$
 \omega_{m+n-2k} = \sum^k_{p=0}(-1)^p {k \choose p} { \prod^{k-1}_{j=0}
(m+n-2k+j+2) \over \prod^{p-1}_{j=0} (n-j) \cdot \prod^{k-p-1}_{j=0}
(m-j) } \cdot f^pv_n \otimes f^{k-p}v_m  .
$$
 Let $B=B^n \otimes B^m$ be the Shapovalov form on the tensor product.

 Consider the rational function
$$
 \Phi (t) = \prod^k_{j=1} t^{-n}_j (t_j-1)^{-m}
\prod_{1 \leq j<i\leq k} (t_i-t_j)^2.
$$
Let $t^0 =(t_1,...,t_k) $ be a critical point of $ \Phi $ such that $t_i
\neq t_j$ for   $i \neq j$.
\proclaim { Theorem}
$$
B(\omega_{n+m-2k}, \omega_{n+m-2k}) \ =\ \det( { \partial^2 \over
\partial t_i \partial t_j } \ln \Phi (t^0)).
$$
\endproclaim

The formula is an analog of the Gaudin and Korepin formulae for the
norm of the Bethe vectors in the theory of quantum integrable models
[Go,K].We prove the formula and its generalizations.

The main results of this work are Theorem (1.2.1),
Theorem (1.3.1),Corollary (2.4.6), and
Theorem (2.5.1).

This work is inspired by [RV] in which the connections between
the quasiclassical asymptotics of solutions of the
Knizhnik-Zamolodchikov equation and the Bethe ansatz vectors for the
Gaudin model are explained.

The author thanks I.Cherednik, N.Reshetikhin and
V.Tarasov for stimulating discussions.

\head  1. Critical Points \endhead

\subhead (1.1) Conjecture \endsubhead

Let $ f_j  :  \Comp ^k \rightarrow \Comp , \ j\ =\ 1,\dots ,N, $
be pairwise different polynomials of degree 1. For every $i$ denote by
$H_i$ the hyperplane in $\Comp^k$ defined by $f_j = 0$. Let $\Cal C
 = \{\ H_j\}^N_{j=1}$ be the configuration of the hyperplanes,
$$\align
&T\ =\ \Comp^k\ -\ \cup^N_{j=1} H_j
\endalign$$
the complement to the union of hyperplanes. Let
$\Lambda  = \{\lambda_j\}^N_{j=1}$
be a collection of complex numbers. Consider the function
$$
\Phi_{\Lambda}\ =\ \prod^N_{j=1} f^{\lambda_j}_j.
\tag1.1.1
$$
$\Phi_{\Lambda}$ is a multivalued holomorphic fumction on $T$. A point
$t \in T$ is critical for $\Phi_{\Lambda}$ if its first derivatives
vanish at $t$. First derivatives vanish at $t$ for all branches of
$\Phi_{\Lambda}$ simultaneously, since the ratio of every two branches
is constant.

Assume that the configuration has a vertex.
\proclaim {Conjecture}
For generic $\Lambda$ all critical points of $\Phi_{\Lambda}$ are
nondegenerate and the number of critical points is equal to
the absolute value of the Euler characteristic of $T$.
\endproclaim
The conjecture is proved below for the case in which all polynomials
$\{ f_j \}$ have real coefficients.
\flushpar{\it Remarks.}

$(a)$ An edge of a configuration is a nonempty intersection of some of
its hyperplanes. A vertex is a zero dimensional edge.

$(b)$ Generic $\Lambda$ means that there exists an algebraic subset
$\Sigma \ \subset \Comp^N$ such that the conjecture is true if $\Lambda
\in \Comp^N  - \Sigma.$

$(c)$ The defining equations for critical points of $\Phi_{\Lambda}$
have the form:
$$
\sum^N_{j=1} \lambda_j{\partial f_j \over \partial t_l}/f_j\ =\ 0,\ \ l\
=\ 1,\dots , k.
$$

$(d)$ According to [OS], the Euler characteristic $\chi (T)$ is defined
combinatorially in terms of the lattice of edges of the configuration:
$$\align
&|\chi (T)|\ =\ |\sum_{E\subset \Cal C} \mu (E)|
\endalign$$
where the sum is over all edges of $\Cal C$, $\mu (E)$ is the
value of the Mobius function of $\Cal C$ at $E$.

$(e)$ There are theorems on Newton polygones in which one considers a
polynomial system of equations depending on parameters. Under certain
conditions on the system, the number of solutions of the system for
generic values of parameters is defined combinatorially in terms of
Newton polygones of equations, see [BKK]. It would be interesting to
find a connection between those theorems and the statement of the
conjecture.

$(f)$ A multidimensional hypergeometric integral is an integral of the
form
$$\align
&\int_{\gamma}\ \Phi_{\Lambda} \ R\ d t_1\wedge \dots \wedge d t_k
\endalign$$
where $R : T \rightarrow \Comp  $ is a rational function and $\gamma
\subset T$ is a suitable cycle. If the polynomials
$\{ f_j\}$ depend on additional parameters, such an integral becomes a
function of additional parameters called a multidimensional
hypergeometric function, see [A,G,V]. Multidimensional hypergeometric
functions satisfy remarkable differential equations. For example, the
Knizhnik-Zamolodchikov equation in conformal field theory is solved in
hypergeometric functions [SV]. In the application to the KZ equation the
exponents $\{ \lambda_j \}$ have the form $\{ \lambda_j=\alpha_j/\kappa
\}$ where $\kappa$ is a parameter of the equation. Studying asymptotics
of solutions of the KZ equation as $\kappa$ tends to 0 leads to studying
critical points of the function $\Phi_{\Lambda}$. This problem motivated
the conjecture.

$(g)$ If the configuration has no vertices, then there exist linear
coordinates
$u_1, \dots , u_k$ in $\Comp^k$ such that all polynomials $ \{ f_j \}$
do not depend on $u_1, \dots , u_r$ for some $r$
and the configuration, cut by $\Cal C$ in $u_1= \dots = u_r=0$,
has a vertex.

\subhead (1.2) Real Configuration \endsubhead
Assume that all polynomials $\ \{f_j \} \ $ have real coefficients :
$$\align
&f_j\ =\  a^0_j + a^1_jt_1 + \dots a^k_jt_k,\ \ j=1, \dots , N,
\endalign $$
and all numbers $\{a^m_j \}$ are real.

Let $T_{\Real} \ =\ T\cap \Real^k.$ Let
 $T_{\Real}\ =\ \cup_{\alpha}D_{\alpha} $
be the decomposition into the union of connected components.
Each component is a convex polytope.

By[BBR], the number of bounded components is given by the formula
$$\align
&\#\ =\  |\sum_{E\subset \Cal C} \mu (E)|
\endalign$$
\proclaim{(1.2.1) Theorem} Let all numbers $\{ \lambda_j\}$ be positive.
Then the union of all critical points of $\Phi_{\Lambda}$
is contained in the union of all bounded components of $T_{\Real}$. Each
bounded component containes exactly one critical point. All critical
points are non-degenerate.
\endproclaim
The theorem implies the conjecture for the case in which all
polynomials $\{f_j\}$ are real.

\demo {Proof} First we formulate a trivial but useful lemma.

Let $t$ be the complex coordinate on $\Comp$, and $r, \phi$
the planar coordinates on $\Comp ,\ t=r\exp (i\phi ).$
Let $v=a{\partial \over \partial \phi}$ be a tangent vector at a point
$t=t^0$ in $\Comp  - 0.$ Denote by $L_v\ln t$ the derivative of $\ln
t$ along $v$.
\proclaim{(1.2.2) Lemma} If $a>0$, then $Re(L_v \ln t ) >0.$
\endproclaim
The theorem is implied by the following three lemmas.
\proclaim{(1.2.3) Lemma} Let $D$ be an unbounded component of
$T_{\Real}$. Then there are no critical points in $D$.
\endproclaim
\demo{Proof} Let $p \in D$. By [BBR] there exists a vector $v \in
\Real^k$ such that the ray $p(s)=p+sv,\ s \in \Real_{\geq 0}$, has no
intersection with the union of hyperplanes of $\Cal C$.
By Lemma (1.2.2) we have
$$
Re(L_v (\ln \Phi_{\Lambda} ) (p)) \ > \ 0.
\tag1.2.4
$$
Hence $p$ is not critical.
\enddemo
\proclaim{(1.2.5) Lemma} Let $p \ \in \ T-\Real^k$. Then $p$ is not
critical.
\endproclaim
\demo{Proof} Let $p=w + iv$ where $w , v \in \Real^k$. The ray
 $p(s)     =p + isv,\ s \in \Real_{\geq 0}$, has no intersection with the
union of hyperplanes of $\Cal C$. By Lemma (1.2.2) we have inequality
(1.2.4). Hence $p$ is not critical.
\enddemo
\proclaim{(1.2.6) Lemma} Let $D$ be a bounded component of $T_{\Real}$.
Then $D$ containes exactly one critical point, and this critical point
is non-degenerate.
\endproclaim
\demo{Proof} We have an equality in $D$:
$$\align
&\ln \Phi\ =\ S \ +\ const
\endalign$$
where $S=\sum \lambda_j\ln |f_j|$. Hence
$\Phi_{\Lambda} $ and $S$ have the same critical points in $D$. We have
$S(p)  \rightarrow  -\infty$ as $p  \rightarrow  \partial D$. So $S$
has a critical point in $D$. The critical point is unique since $S$ is
convex. The critical point is non-degenerate because the matrix of
second derivatives of $S$ is positive definite.
\enddemo
\enddemo
\subhead (1.3) Example \endsubhead
Let
$$\align
&\Phi (t)\ =\ \Phi (t;\alpha,\beta ,\gamma)\ =\ \prod^k_{j=1}
t^{\alpha}_j (1-t_j)^{\beta} \prod_{1\leq j<i\leq k} (t_i-t_j)^{2\gamma
},
\endalign$$
where $\alpha , \beta ,\gamma $ are complex parameters. We describe
critical points of $\Phi$.

The critical set of $\Phi$ is invariant with respect to the group of
permutations of coordinates. By Theorem (1.2.1) the number of critical
points is not greater than $k!$.

Let $\lambda_1=t_1+ \dots + t_k, \lambda_2 = \sum t_it_j, \dots ,
\lambda_k = t_1 \cdot \dots \cdot t_k$ be
the standard symmetric functions. Let
$\mu_1=(1-t_1)+ \dots +(1-t_k),\ \mu_2=\sum (1-t_i)(1-t_j), \dots ,
\mu_k=(1-t_1) \cdot \dots \cdot (1-t_k), \
\delta = \prod_{1\leq i<j\leq k} (t_i-t_j)^2.$
\proclaim{(1.3.1) Theorem} If $(t_1,\dots ,t_k)$ is a critical point of
$\Phi$, then
$$
\lambda_l\ =\  {k \choose l}  \prod^l_{j=1} { \alpha +
(k-j)\gamma \over \alpha + \beta +(2k-j-1)\gamma},
\tag1.3.2
$$
$$\align
&\mu_l\ =\  {k \choose l}  \prod^l_{j=1} { \beta +
(k-j)\gamma \over \alpha + \beta +(2k-j-1)\gamma},
\endalign
$$
for all $l$.
\endproclaim
\demo{Proof}  The defining equations for critical points are
$$\align
&{\alpha \over t_i} + {\beta \over t_i-1} +\sum_{j\neq i} {2\gamma \over
t_i-t_j}\ =\ 0,\ \ i=1,\dots ,k.
\endalign$$
Multiplying the $i$-th equation by $t_i/t_1\cdot \dots \cdot
t_k$ and taking the sum
of the equations we get
$$\align
&{k(\alpha +(k-1)\gamma) \over t_1\cdot \dots \cdot
t_k} +\beta ({1 \over t_1\cdot \dots \cdot
t_{k-1}(t_k-1)} + \dots + {1 \over(t_1-1)t_2 \cdot \dots \cdot t_k})\ =\ 0.
\endalign$$
Similarly, for every $p=0, \dots , k-1,$ we get
$$\align
&(k-p)(\alpha +(k-p-1)\gamma ) \sum_{1 \leq i_1 < \dots < i_{p} \leq
k} {1 \over (t_{i_1}-1)} \dots {1 \over (t_{i_{p}}-1)} \prod_{j \not\in
(i_1, \dots ,i_{p})} {1 \over t_j} \
 + \
\\&(p+1)(\beta +p\gamma ) \sum_{1 \leq i_1 < \dots < i_{p+1} \leq
k} {1 \over (t_{i_1}-1)} \dots {1 \over (t_{i_{p+1}}-1)} \prod_{j \not\in
(i_1, \dots ,i_{p+1})} {1 \over t_j}\  =\ 0.
\endalign$$
This system of equations implies
$$ \multline
(k-p)(\alpha +(k-p-1)\gamma ) \sum^k_{j=0} (-1)^j  {k-j \choose
 p }  \lambda_{k-j} +
\\
(p+1)(\beta +p\gamma ) \sum^k_{j=0} (-1)^j  {k-j \choose p+1}
 \lambda_{k-j} \ =\ 0
\endmultline
\tag1.3.3
$$
 for $p=0, \dots ,k-1.$ Here $\lambda_0=1.$
\proclaim {(1.3.4) Lemma}
System (1.3.3) is equivalent to the system:
$$\align
&(p+1)(\alpha +p\gamma)\lambda_{k-p-1} \ =\
(k-p)(\alpha +\beta +(k+p-1)\gamma)\lambda_{k-p}
\endalign$$
for $p=0, \dots ,k-1.$
\endproclaim
Lemma (1.3.4) is proved by induction on $p$. Lemma (1.3.4) proves
Theorem (1.3.1).
\enddemo
Set
$$\align
&A_{k-p,p}(t;\alpha ,\beta ,\gamma )\ = \
\sum_{1 \leq i_1 < \dots < i_{p} \leq
k} {1 \over (t_{i_1}-1)} \dots {1 \over (t_{i_{p}}-1)} \prod_{j \not\in
(i_1, \dots ,i_{p})} {1 \over t_j}
\endalign$$
for $p=0, \dots ,k.$
\proclaim{(1.3.5) Lemma}
If $(t^0_1, \dots ,t^0_k)$ is a critical point of the function
$\Phi (t;\alpha ,\beta ,\gamma ) $ , then
$$\align
&A_{k-p,p}(t^0;\alpha ,\beta ,\gamma ) = (-1)^p  {k \choose p}
{ \prod^k_{j=1} ( \alpha + \beta  +
(2k-j-1)\gamma ) \over
 \prod^{k-p-1}_{j=0} (\alpha + j\gamma )
 \cdot \prod^{p-1}_{j=0} (\beta +j\gamma) } .
\endalign$$
\endproclaim
The vector
$A_k\ =\ (A_{k,0}, \dots ,A_{0,k} )$
has the following interpretation.

Consider the Lie algebra $sl_2$ with the standard generators $e,\ f,\
h. $
For $\alpha \in \Comp $ let $V_{\alpha}$ be an $sl_2$ module with highest
weight $\alpha$, that is, the module $V_{\alpha}$ is generated by a
vector $v_{\alpha}$ such that $ev_{\alpha}=0$ and $hv_{\alpha}=\alpha
v_{\alpha}.$ For $\alpha , \beta \in \Comp$ consider the vector
$$\align
&F_k(\alpha , \beta ) \ =\ A_{k,0}(t^0;\alpha ,\beta ,-1 )
f^kv_{\alpha}
\otimes v_{\beta} + \dots +
A_{0,k}(t^0;\alpha ,\beta ,-1 )
v_{\alpha}
\otimes f^k v_{\beta}
\endalign$$
of the tensor product $V_{\alpha} \otimes V_{\beta}.$ We have
$hF_k \ =\ (\alpha +\beta -2k)F_k.$

\proclaim {Corollary of Lemma (1.3.5)} The vector $F_k(\alpha ,\beta )$
is a singular vector:$\ eF_k  = 0.$
\endproclaim

Explanations of this fact see in [RV] and in Section 2.

The Shapovalov form on $V_{\alpha}$
 is the unique symmetric bilinear form $B_{\alpha}$
 defined by the conditions:
$$\align
&B_{\alpha}(v_{\alpha},v_{\alpha})  = 1,\  \  \
B_{\alpha}(fx,y) = B_{\alpha}(x,ey),
\endalign$$
for all $x,y \in V_{\alpha}.$ Consider the bilinear form $B =
B_{\alpha}
\otimes B_{\beta}$ on
$V_{\alpha}
\otimes V_{\beta}$ .
\proclaim{(1.3.6) Lemma}
We have
$$\align
&B(F(\alpha ,\beta ), F(\alpha ,\beta )) \ =\  k! \prod^{k-1}_{l=0}
{(\alpha + \beta -2k+l+2)^3 \over (\alpha -l) (\beta -l)}.
\endalign$$
\endproclaim
The proof easily follows from the formula
$$\align
&\prod^{k-1}_{l=0} (\alpha + \beta -2k+l+2) \ =\  \sum^k_{p=0}
 {k \choose p}  \prod^{p-1}_{l=0} (\alpha -k+l+1)
\prod^{k-p-1}_{l=0} (\beta -k+l+1),
\endalign$$
and the formula can be proved by induction.

Denote the number $B(F(\alpha ,\beta ), F(\alpha ,\beta ))$ by $b(\alpha
, \beta ;k).$

\subhead (1.4) Asymptotics of Selberg Integral
\endsubhead

The Selberg formula,
$$\multline
k! \int_{\Delta} \prod^k_{j=1} t^{\alpha}_j
(1-t_j)^{\beta} \prod_{1
\leq i<j \leq k} (t_i-t_j )^{2\gamma } d t_1 \wedge \dots \wedge
 d t_k \ =
\\
\prod^{k-1}_{l=0}{ \Gamma (\alpha + l\gamma +1) \Gamma (\beta +l\gamma
+1) \Gamma ((l+1)\gamma +1) \over \Gamma (\alpha +\beta +(2k-l-2)\gamma
 +2) \Gamma (\gamma +1)}
\endmultline\tag1.4.1
$$
where $\Delta = \{ t \in \Real^k \ |\ 0 < t_1 < \dots <t_k<1\} $
(see [A,As,M,S]), has beautiful applications, in particular in conformal
field theory [DF].

Assume that $\alpha = a/\kappa , \beta = b/\kappa , \gamma =c/\kappa$
where $a,b,c,\kappa$ are positive numbers and $\kappa$ tends to zero.
We compute asymptotics of both sides of formula (1.4.1).

The method of steepest descent gives the following asymptotics for the
left hand side of (1.4.1):
$$\align
&lhs \ \sim \ k!\cdot (2\pi \kappa )^{k/2}
\cdot \Phi (t^0;a /\kappa , b/\kappa , c/ \kappa ) \cdot
\Hess (-S(t^0;a,b,c))^{-1/2}
\endalign$$
where
$$\align
S(t;a,b,c)) \ =&\ \kappa \ln
\Phi (t;a /\kappa , b/\kappa , c/ \kappa ) \ =
\\&  \sum^k_{j=1}
(a\ln t_j + b \ln (1-t_j)) + \sum_{1 \leq i<j \leq k}
2c \ln (t_i-t_j) ,
\endalign$$
$t^0$ is the critical point of $S$ in $\Delta$,
$$\align
&\Hess (-S) \ =\ \det(-{\partial ^2 S \over \partial t_i \partial t_j }
),
\endalign$$
$$\align
-{\partial ^2 S \over \partial t^2_i } \ =\ a{1 \over t^2_i}+b{1 \over
(t_i -1)^2} + 2c\sum_{j \neq i} {1 \over (t_i-t_j)^2},\ \ \
\ \ \ -{\partial ^2 S \over \partial t_i \partial t_j} \ =\ -2c{1 \over
(t_i-t_j)^2}.
\endalign$$
The symmetric functions of coordinates $t^0_1, \dots ,t^0_k$ are
given by  formula (1.3.2) in which $\alpha ,\beta ,\gamma $ must be
replaced by $a,b,c,$ respectively.

Let $\lambda _k = t^0_1 \cdot \dots \cdot t^0_k ,\
 \mu _k = (1-t^0_1)\cdot \dots \cdot (1-t^0_k),
\ \delta = \prod_{i<j} (t^0_i-t^0_j)^2.$ Then
$$\align
&lhs \ \sim k!\cdot (2\pi \kappa )^{k/2} \cdot
\lambda ^{a/\kappa}_k \cdot \mu ^{b/\kappa
}_k \cdot \delta ^{c/\kappa}
\Hess (-S(t^0;a,b,c))^{-1/2}.
\endalign$$
Asymptotics of the right hand side of (1.4.1) can be computed by the
Stirling formula. Comparing both asymptotics we get the following
formulae:
$$
\delta \ =\ \prod^{k-1}_{l=0} {(l+1)^{l+1}c^l(a+lc)^l(b+lc)^l
\over (a+b+(2k-l-2)c)^{2k-l-2}},
\tag1.4.2
$$
$$
\Hess(-S(t^0;a,b,c)) \ =\ k! \prod^{k-1}_{l=0}
{(a+b+(2k-l-2)c)^3 \over (a+lc)(b+lc)}.
\tag1.4.3
$$Formula (1.4.3) is an example of the series of rather surprising
formulae in which the determinant of a bilinear form is the product of
very simple factors \newline
[V1-3,L,LS,SV,BV]. The list of such examples
includes formula (1.4.1), the Legandre equation, the Vandermonde
determinant, and many others.

Comparing formula (1.4.3) and Lemma (1.3.6) we get
\proclaim{(1.4.4) Theorem}
Let $F(\alpha ,\beta ) \in V_{\alpha} \otimes V_{\beta} $ be the vector
defined in Section (1.3), $B$ the Shapovalov form on
$V_{\alpha} \otimes V_{\beta} $ . Then
$$\align
&B(F(\alpha ,\beta ),F(\alpha ,\beta )) \ =\  \Hess (-S(t^0,\alpha ,\beta
,-1)).\endalign$$
\endproclaim
This theorem implies the theorem formulated in the introduction.

\head 2. Families of Bases of Singular Vectors. \endhead
\subhead (2.1) KZ equation \endsubhead
Consider the Lie algebra $\frak g=sl_2$ with the generators $e,f,h.$
Let $\frak h$ be the Cartan subalgebra of $\frak g$ generated by $h$,
let $\alpha \in \frak h^* $ be
the simple root, $(,)$ the bilinear form on
$\frak h ^*$  such that $(\alpha ,\alpha )=2.$

Denote by $\Omega$ the element
$h \otimes h/2+e\otimes f+ f\otimes e \in \frak g \otimes \frak g$
corresponding to the Killing form.

Let $V_1, \dots ,V_n$ be $\frak g$ modules, $V=V_1 \otimes \dots
\otimes V_n.$ For $i<j$ let $\Omega _{i,j}$ be the linear operator on
$V$  acting as $\Omega$ on $V_i \otimes V_j$ and as the identity
operator on other factors.

The {\it Knizhnik-Zamolodchikov equation (KZ)}
 on an $\ V$-valued function
\newline $ \Psi
(z_1,  \dots  , z_n) $ is the system of equations
$$\align
&\kappa{\partial \Psi \over \partial z_i} \ =\ H_i\Psi,\ \ \ i=1, \dots
,n.
\endalign$$
where $\kappa$ is a parameter of the equation and
$$\align
&H_i \ =\ \sum_{j \neq i} {\Omega _{i,j} \over z_i-z_j}.
\endalign$$
For $\Lambda _1, \dots ,\Lambda _n \in \frak h ^*$ let $ V_1, \dots
 ,V_n$ be highest weight $\frak g$ modules with highest weights $\Lambda
_1, \dots , \Lambda _n$ and highest weight vectors $v_1, \dots v_n,$
respectively.

For a nonnegative integer $k$ let
$$\align
&(V)_k = \{ v\in (V)_k \ | \ hv=(\sum^n_{i=1} \Lambda _i-k\alpha ,\alpha
)v\}
\endalign$$
be the weight space and
$$\align
&Sing(V)_k\ =\ \{v \in (V)_k \ |\ ev=0\}
\endalign$$
the subspace of singular vectors of weight $k$.

The KZ equation preserves the subspace of singular vectors.

Let $B_i$ be the Shapovalov form on $V_i$. Denote by $B$ the bilinear
form
$B_1 \otimes \dots \otimes B_n$ on $V$. If $V_1,\dots ,V_n$ are
irreducible, then $B$ is nondegenerate. It is easy to see that the
operators $H_1,\dots ,H_n$ are symmetric:
$$\align
&B(H_ix,y)=B(x,H_iy)
\endalign$$
 for all $i$ and all $x,y \in V$.

\subhead (2.2) Integral Representations \endsubhead

There is an integral representation for solutions of the KZ equation
with values in $Sing(V)_k$ [SV].

Set
$$\align
&\Phi (t,z) \ =\ \prod_{1\leq  i<j \leq n} (z_i-z_j)^{(\Lambda _i,\Lambda
_j)/\kappa } \prod_{1 \leq j<i \leq k} (t_i-t_j)^{2/\kappa }
\prod_{i,j} (t_i-z_j)^{-(\Lambda _j,\alpha )/\kappa }.
\endalign$$

{\it A monomial} of weight $k$ is an element of $(V)_k$ of the form
$$\align
&f_I \ =\ f^{i_1}v_1\otimes \dots \otimes f^{i_n}v_n,
\endalign$$
here $I=(i_1, \dots ,i_n),\ i_1+ \dots +i_n=k.$

For a monomial $f_I$, define a differential $k$-form in $t$
and $z$:
$$\align
&\eta (f_I) \ =\ A_I(t,z)\ d t_1 \wedge \dots \wedge d t_k,
\\
& A_I(t,z)\ =\ \sum_{\sigma \in S(k,i_1, \dots ,i_n)}
\prod^k_{i=1}{1 \over (t_i- z_{\sigma (i)})}.
\endalign$$
The sum is over the set $S(k,i_1, \dots ,i_n)$ of maps $\sigma$ from
$\{1, \dots ,k\}$
 to $\{1, \dots , n \}$, such that for all $m$ the cardinality of
$\sigma ^{-1}(m)$ is $i_m$.

Let $\Cal C = \Cal C (z_1, \dots ,z_n)$ be the configuration of
hyperplanes
$$\align
&t_i=t_j,\  1\leq i<j \leq k, \ \ \ \ \ \
 t_i=z_j,\ i=1, \dots , k,\
j=1, \dots ,n,
\endalign$$
 in $\Comp ^k$. Let $T(z_1, \dots ,z_n)$ be the
complement to the union of hyperplanes of $\Cal C$ in $\Comp ^k$.

Assume that $z_1, \dots ,z_n$ are pairwise different real numbers. Let
$D(z_1, \dots ,z_n)$ be a bounded component of $T(z_1, \dots ,z_n)
\cap \Real ^k$ continuously depending on $z$.
\proclaim {(2.2.1) Theorem [SV]} The function
$$\align
&\Psi _D(z) \ =\ \sum_{f_I \in V_k} \int _{D(z)} \Phi (t,z)\cdot
 \eta (f_I)
\cdot f_I
\endalign$$
takes values in $Sing(V)_k$ and satifies the KZ equation.
\endproclaim
{\it Remark.} If the integrals diverge, then their value must be taken
in the sence of analytic continuation with repect to parametars $\Lambda
_1, \dots ,\Lambda _n,\kappa $. If $z_1, \dots z_n$ are not real, then
$D(z)$ must be replaced by a $k$-cycle in $T(z)$ with coefficients in a
suitable local system, see [SV, V5].

\subhead (2.3) Basis of solutions \endsubhead
For $i=1, \dots ,n$, let $V_i$ be the Verma module with highest weight
$\Lambda _i$. This means that $V_i$ is an infinite dimensional module
generated by a vector $v_i$ such that $hv_i=(\Lambda _i, \alpha )v_i$
and $ev_i=0$.

Consider the set of solutions $\{\Psi _D\}$ where $D$ ranges over all
bouded components of $T-\Real ^k$. According to [SV], $\{ \Psi _D \}$
generate all solutions of the KZ equation with values in $Sing(V)_k$ for
generic $\Lambda _1, \dots ,\Lambda _n, \kappa$. Below we'll give a
formula for a suitable determinant which will make the above statement
more explicit.

Assume that $z_1<z_2< \dots <z_n.$ We say that a bounded component $D(z)
\in T(z)-\Real ^k$ is {\it admissible}
 if it lies in the cone $z_1<t_1<t_2<
\dots <t_k.$ Admissible components have the form
$$\align
D(z) \ =\ \{t \in \Real ^k \ |\ z_1<t_1< \dots <t_{k_2}&<z_2<
\\
&\dots <z_{n-1}< t_{k_{n-1}+1}< \dots < t_k<z_n \}
\endalign$$
where $0\leq k_2 \leq \dots \leq k_n=k.$

A monomial $f_I \in (V)_k$ for
$I=(0,i_2, \dots ,i_n),\ i_2= \dots +i_n=k,$
will be called {\it admissible}.

The number of admissble components is equal to the number of admissible
monomials. Denote this number by $N$.

Set
$$\align
&\overline{\Phi} (t,z) \ =\ \prod_{1 \leq j<i \leq k}
(t_i-t_j)^{2/\kappa}\prod^k_{i=1}\prod^n_{j=1}(t_i-z_j)^{-(\Lambda
_j,\alpha )/\kappa }.
\endalign$$
Consider the determinant $\det (\int_{D(z)} \overline{\Phi}\cdot
 \eta (f_I))
$ where $D(z)$ ranges over all admissible components and $f_I$ ranges
over all admissible monomials. We'll give a formula for this
determinant.

For any admissible domain $D$ and for any function $g$ of the form
$$
\multline
(t_i-t_j)^{2/\kappa},\ \ \  1\leq i<j \leq k,
\\
(t_i-z_j)^{-(\Lambda _j,\alpha )/\kappa },\ \ \
i=1,\dots ,k,\ j=1, \dots ,n,
\endmultline
\tag2.3.1
$$
fix a branch of $g$ over $D$. This choice determines a branch of
$\overline{\Phi}$ and, hence, a branch of the determinant.

For any such $g$ and for any admissible $D$, let $t$ be a point of the
closure of $D$ which is the most remote point from the hyperplane
of singularities of $g$. The number $g(t)$ will be called the {\it
extreme
value} of $g$ on $D$ and will be denoted by $c(D,g)$.

For every $g$ and $D$, the number $c(D,g)$ is equal to
$(z_l-z_m)^{-(\Lambda _l,\alpha )/\kappa}$, or \newline
$(z_l-z_m)^{-(\Lambda _m,\alpha )/\kappa}$, or
$(z_l-z_m)^{2/\kappa}$ , for suitable $l$ and $m$.

For any admissible monomial $f_I$, set
$$\align
&b(f_I) \ =\ \prod^n_{l=2} i_l! \kappa ^{-i_l} \prod^{i_l-1}_{j=0}
(-(\Lambda _l,\alpha )+j).
\endalign$$

\proclaim {(2.3.2) Theorem [V3]} We have
$$\multline
\det (\int_{D(z)} \overline{\Phi}\cdot  \eta (f_I)) \ =\ \pm \prod _{D(z),g}
c(D(z),g)\cdot \prod _{f_I} b(f_I)^{-1} \cdot
\\
 \prod^{k-1}_{i=0} \Biggl( {\Gamma ({i+1 \over \kappa } +1)^{n-1}
\prod^n_{l=1} \Gamma ({-(\Lambda _l,\alpha ) +i \over \kappa } +1)
\over \Gamma ({1 \over \kappa } +1)^{n-1} \Gamma (
\sum^n_{l=1}{-( \Lambda _l, \alpha ) \over \kappa } + {2k-i-2 \over \kappa
}+1)} \Biggr) ^{p_i}
\endmultline
\tag2.3.3
$$
where $p_i={n+k-i-3 \choose k-i-1}$, the first product is
 over all admissible domains and over all functions described in
(2.3.1), the second product is  over all admissible monomials.
\endproclaim

The sign $\pm$ in the formula depends on the choice of the orientation
of domains, see [V1, V2].

It is easy to see that
$$\align
&\dim Sing(V)_k \ =\ N
\endalign$$
if $(\Lambda _1, \alpha )\ \neq \ 0,1, \dots , k-1.$

\proclaim {(2.3.4) Corollary} Assume that $\Lambda _1, \dots , \Lambda _n
, \kappa$ are such that
$(\Lambda _1, \alpha )\ \neq \ 0, 1,  \dots  ,$
\newline $  k - 1,$
and the right hand side of (2.3.3) is well-defined and not equal to
zero. Then $\{ \Psi _D \}$ form a basis of solutions of the KZ equation
with values in $Sing(V)_k$, where $D$ ranges over admissible components.
\endproclaim

\subhead (2.4) Quasiclassical Asymptotics \endsubhead

Assume that $\Lambda _1, \dots ,\Lambda _n$ are such that
$-(\Lambda _1,\alpha ), \dots , -(\Lambda _n, \alpha )$ are positive.
Assume that $\kappa$ is positive and tends to zero. We'll compute
asymptotics of the basis $\{ \Psi _D \}$ , see Corollary (2.3.4).

For an admissible domain $D(z)$, let $t_D(z)$ be the unique critical
point of $\Phi$ in $D(z))$, see Theorem (1.2.1). The point $t_D(z)$
depends on $z,\Lambda _1,\dots ,\Lambda _n$ and does not depend
on $\kappa$.

By the method of steepest descent we have an asymptotic expansion
$$\multline
\Psi _D(z) =
\\
(2\pi \kappa )^{k/2} \cdot \Phi (t_D(z),z) \cdot
\Hess_t(-S(t_D(z),z))^{-1/2} \cdot (F(t_D(z),z) + \Cal O (\kappa)),
\endmultline
\tag2.4.1
$$
where $S=\kappa \ln \Phi,$
$$\align
&F(t,z) \ =\ \sum_{f_I \in (V)_k} A_I(t,z)\cdot f_I,
\endalign
$$
the sum is over all monomials in $(V)_k$.
\proclaim {(2.4.2) Theorem [RV]} The vector $F(t_D(z),z)$ lies in
$Sing(V)_k$. For any $l=1, \dots , n,$ the vector $F(t_D(z),z)$ is
an eigenvector of the operator $H_l(z)$ with eigenvalue
${\partial S \over \partial z_l} (t_D(z),z) $. Moreover,
$$\align
&B(F(t_D(z),z),F(t_D(z),z)) \ =\ const \Hess _t(S(t_D(z),z))
\endalign$$
where $B$ is the Shapovalov form defined in Section (2.1), and $const $
does not depend on $z$.
\endproclaim

{\it Remark.} Theorem (1.4.4) states that $const = 1$ if $n=2$.

\proclaim {(2.4.3) Theorem} We have the following two formulae.
$$\multline
\prod _D  \overline {\Phi} (t_D(z),z)  = \prod_{D,g} c(D,g) \cdot
\\
\cdot \prod^{k-1}_{i=0} \Biggl( {(i+1)^{{(n-1)(i+1)\over
 \kappa }}\prod^n_{l=1} (-(\Lambda _l, \alpha ) +i)^{{-
(\Lambda _l, \alpha) +i \over \kappa}}
\over (-\sum^n_{l=1} (\Lambda _l, \alpha ) +2k-i-2)^
{ {-\sum^n_{l=1} (\Lambda _l, \alpha ) +2k-i-2 \over \kappa}}}
\Biggr) ^{p_i}.
\endmultline
\tag2.4.4
$$

Here the first product is over admissible domains, the second product
and $p_i$ are explained in Theorem (2.3.2).
$$\multline
\det (A_I(t_D(z),z)) = \pm \prod _D \Hess _t(-S(t_D(z),z))^{1/2}
\cdot
\\
\cdot \prod _{f_I} \overline {b} (f_I)^{-1} \prod ^{k-1}_{i=0}
\Bigl({(i+1)^{n+1}\prod^n_{l=1}
(-(\Lambda _l, \alpha ) +i)
\over
-\sum^n_{l=1} (\Lambda _l, \alpha ) +2k-i-2}
\Bigr) ^{p_i/2}.
\endmultline
\tag2.4.5
$$
Here the rows (columns) of the determinant are numerated by admissible
domains (admissible monomials). The first (second) product is over
admisible domains (admissible monomials), and
$$\align
&\overline {b} (f_I) \ =\ \prod ^n_{l=2} i_l!\prod ^{i_l-1}_{j=0}
(-(\Lambda _l, \alpha ) +j).
\endalign$$
\endproclaim
To prove the theorem it is enough to write an asymptotic espansion for
the
$rhs$ of (2.3.3) by the Stirling formula, for the $lhs$
 by formula (2.4.1),
and then to compare the corresponding terms.
\proclaim{(2.4.6) Corollary} Vectors $\{ F(t_D(z),z)\}$ form a basis
in $Sing(V)_k$ if \newline $  -(\Lambda _1, \alpha ),  \dots
 ,  -(\Lambda _n,  \alpha)$ are
positive and $z_1, \dots , z_n$ are real paiwise different numbers.
\endproclaim
{\it Remarks.}

1. There are integral formulae for solutions of the KZ equation
associated with an arbitrary Kac-Moody algebra [SV]. Theorem(2.4.2)
holds in this more general context [RV]. A conjectural analog of Theorem
(2.3.3) is formulated in [V1]. It is plausible that the conjectural
determinant formula in [V1] would imply that the eigenvectors given by
the first tems of asymptotic expansions of solutions of the KZ equation
for a Kac-Moody algebra generate a basis of the corresponding space of
singular vectors.

2. It is plausible that there are analogs of formulae of Theorem (2.4.3)
for an arbitrary configuration of hyperplanes, and these analogs are
corollaries of the conjectural determinant formula in [V1].

\subhead (2.5) Norms of Eigenvectors \endsubhead
\proclaim{(2.5.1) Theorem} Under assumptions of Theorem (2.4.3) we have
$$\align
&B(F(t_D(z),z),F(t_D(z),z)) \ =\  \Hess _t (S(t_D(z),z)).
\endalign$$
\endproclaim
\proclaim{Corollary} For arbitrary $\Lambda _1,  \dots ,
 \Lambda _n$ and
for arbitrary nondegenerate critical point $t=t(z)$ of the function
$\Phi (t,z)$, we have
$$\align
&B(F(t(z),z),F(t(z),z)) \ =\  \Hess _t (S(t(z),z)).
\endalign$$
\endproclaim
\demo {Proof of the Theorem} Let $W_i,\ i=1,2,$ be $sl_2$ modules.
Let $w_i \in W_i$ be a singular vector of weight $m_i \in \Comp$, that
is, $ew_i=0$ and $hw_i=m_iw_i.$
For a nonnegative integer $l$, the vector
$$\align
&(w_1,w_2)_l:=
\sum^l_{p=0} (-1)^p {l \choose p}
{\prod^{l-1}_{j=0} (m_1 +m_2+j+2-2l) \over
\prod^{p-1}_{j=0}(m_1-j) \cdot \prod^{l-p-1}_{j=0}(m_2-j)}
f^pw_1 \otimes f^{l-p} w_2
\endalign$$
is a singular vector in $W_1 \otimes W_2$ of weight $m_1 +m_2-2l$.

Let $V_1, \dots ,V_n$ be $sl_2$ modules with highest vectors $v_1, \dots
,v_n$
and highest weights $\Lambda _1, \dots ,
 \Lambda _n$, respectively. Set
$m_i=(\Lambda _i, \alpha ).$ For any sequence of non-negative integers
$I=(i_2, \dots , i_n),\ i_2 + \dots + i_n=k,$ set
$$
v_I \ =\ ( \dots ((v_1,v_2)_{i_2},v_3)_{i_3}, \dots , v_n)_{i_n}.
\tag2.5.2
$$
The vector
$v_I$ is a singular vector in $V=V_1 \otimes \dots \otimes V_n$ of
weight $m_1 + \dots + $ \newline $ +m_n -2k.$
\proclaim {(2.5.3) Lemma} Let $B=B_1 \otimes \dots \otimes B_n$ be the
Shapovalov form on $V$. Then
$$\align
&B(v_I,v_I) \ =\ \prod^n_{l=2} b(m_1+ \dots +m_{l-1} -2
(i_2 + \dots + i_{l-1}),m_l;i_l)
\endalign$$
where $b(\alpha, \beta ; i)$ is defined in Section (1.3).
\endproclaim
The lemma easily follows from Lemma (1.3.6).

For any $l=2, \dots , n$, set
$$\align
& G_l \ =\ \sum_{i<l} \Omega _{i,l}.
\endalign$$
\proclaim {(2.5.4) Lemma [RV]} For every $l$ and  $I=(i_2, \dots
,i_n)$, the iterated vector $v_I$ is an eigenvector of $G_l$ with
eigenvalue
$$\align
&\lambda (m_1+ \dots +m_{l-1} -2(i_2 + \dots +i_{l-1}), m_l;i_l)
\endalign$$
where $\lambda (a,b;i)={1 \over 2}ab - i(a+b) +i(i-1).$
\endproclaim
Under assumtions of Theorem (2.4.3) assume that $z_j=s^j,\ j=1, \dots ,
n,$ and $s$ tends to $+ \infty$.

For any $I=(i_2, \dots ,i_n), \ i_2 + \dots +i_n=k,$ let
$$\align
D_I(z)=\{ t \in \Real ^k\ |\ z_1<t_1< \dots < t_{i_2}&<z_2< \dots
\\
&<z_{n-1}<t_{i_2+ \dots +i_{n-1}+1}< \dots <t_k < z_n\}
\endalign$$
be the admissible domain corresponding to this sequence. Let $t_I(z)$ be
the critical point of $\Phi$ in $D_I(z)$. For any $l=2, \dots ,n$,
introduce the function
$$
S_l \ =\ -\sum^{i_l}_{j=1} (\alpha _l \ln t_j + m_l \ln (t_j-1)) +
2\sum_{1 \leq i<j \leq i_l} \ln (t_i-t_j)
\tag2.5.5
$$
where $\alpha _l=m_1+ \dots + m_{l-1} -2(i_2 + \dots +i_{l-1}).$
Let $t^l$ be a critical point of $S_l$.
\proclaim {(2.5.6) Lemma } We have

1.
$$\align
&F(t_I(z),z) \ =\ s^{d(I)}(v_I + \Cal O (s^{-1}))
\endalign$$
where $d(I)$ is some integer, see [RV].

2.
$$\align
&\Hess (-S(t_I(z),z)) \ =\ s^{c(I)}( \prod ^n_{j=2}
 \Hess (-S_j (t^j)) + \Cal O (s^{-1}))
\endalign$$
where $c(I)$ is some integer.

3. For any $l=2, \dots , n$, the operator $H_l(z) =\sum_{j \neq l}
\Omega _{j,l}/(z_j-z_l)$ has the following asymptotics:
$$\align
&H_l(z(s)) \ =\ s^{-l} (\Omega _{1,l} + \dots + \Omega_{l-1,l}
 + \Cal O (s^{-1})).
\endalign$$
\endproclaim
\demo{Proof} Make a change of variables: $t_j = s^lu_j$ if
$i_2+ \dots +i_{l-1} <j<i_2 + \dots +i_l.$ Then we have
$$
S(t(u)) \ =\ A \ln s +S_2(u)+ \dots + S_n(u) + \Cal O (s^{-1})
\tag2.5.7
$$
for some number $A$. This formula and the explicit formula for $F(t,z)$
imply the lemma.
\enddemo
By Theorem (1.4.4) and Lemma (2.5.3) we have
$$\align
&B(v_I,v_I) \ =\ \prod^n_{j=2} \Hess (S_j(t^j)).
\endalign$$

Now Theorem (2.4.3) implies equality $2d(I)=c(I)$ and Theorem (2.5.1).
\enddemo
{\it Remark.} Using part 3 of Lemma (2.5.6) and Lemma (2.5.4) we can
compute asymptotics of eigenvalues of the operators $H_1(z), \dots ,
H_n(z)$ on vectors \newline
$\{F(t_D(z),z)\}$ for $z_1<<z_2<< \dots <<z_n$.
This computation shows that the eigenvalues separate the vectors. This
means that the vectors are pairwise orthogonal with respect to the
Shapovalov form [RV].

\heading {\bf REFERENCES} \endheading

\item{[A1]} K.Aomoto, {\it On the structure of integrals of power
product of linear functions,}
Sc. Papers of the College of General Education, Tokyo Univ.{\bf 27}
(1977), 49-61.
\item{[A2]} K.Aomoto, {\it On the complex Selberg integral,}
Q. J. Math. Oxford
{\bf 38} (1987), 385-399.
\item{[As]} R.S.Askey, {\it Some basic hypergeometric extensions of
integrals of Selberg and Andrews,} SIAM J. Math. {\bf 11} (1980),
938-951.
\item{[BBR]} M.Barnabei, A.Brini, G.-C. Rota, {\it Theory of Mobius
functions,} Uspehi Mat. Nauk {\bf 41} (1986), no. 3, 113-157.
\item{[BBK]} D.Bernstein, A.Kushnirenko, A.Khovansky, {\it Newton
polytopes,} Uspehi Mat.\newline Nauk {\bf 31} (1976), no. 3, 201-202.
\item{[BV]} T.Brylawski, A.Varchenko, {\it
Determinant formula for the bilinear form of a matroid,} Preprint, 1993.
\item{[DF1]} V.Dotsenko, V.Fateev, {\it Conformal algebra and
multipoint correlation functions in 2D statistical models,} Nucl. Phys.
{\bf B240} (1984), 312-348.
\item{[DF2]} V.Dotsenko, V.Fateev, {\it Four-point correlation
functions and the operator algebra in 2D conformal invariant theories
with central charge
$c \leq 1$, } Nucl. Phys. {\bf B251} (1985),
691-734.
\item{[G]} I.M.Gelfand, {\it General theory of
hypergeometric functions ,} Dokl.  Akad. Nauk. SSSR {\bf 288}
 (1986), 573-576.
\item{[Go]} M.Gaudin, {\it Diagonalization d'une classe d'hamiltoniens
de spin,}
\newline Journ. de Physique {\bf 37} (1976), no. 10, 1087-1098.
\item{[K]} V.Korepin, {\it Calculation of Norms of Bethe Wave Functions}
, Comm. Math. Phys. {\bf 86} (1982), 391-418.
\item{[L]} F.Loeser, {\it Arrangements d'hyperplans et sommes de
Gauss,} Ann.  Scient. Ec. Norm. Sup. {\bf 24} (1991), 379-400.
\item{[LS]} F.Loeser, C.Sabbah, {\it Equations aux differences finies
et determinants d'integrales de fonctions multiformes,} Comment. Math.
Helv. {\bf 66} (1991), 345-361.
\item{[M]} M.Mehta, {\it Random Matrices,} Acad. Press, 1991.
\item{[OS]} P.Orlik, L.Solomon, {\it Combinatorics and Topology of
Complements of Hyperplanes,} Invent. math {\bf 56} (1980), 167-189.
\item{[RV]} N.Reshetikhin, A.Varchenko, {\it Quasiclassical Asymtotics
of Solutions of the KZ Equations,} Preprint, 1993.
\item{[SV]} V.Schechtman, A.Varchenko, {\it Arrangements of
hyperplanes and Lie algebra homology,} Invent. math. {\bf 106} (1991),
139-194.
\item{[S]} A.Selberg, {\it Bemerkninger om et multiplet integral,}
Norsk. Mat.  Tidsskr. {\bf 26} (1944), 71-78.
\item{[V1]} A.Varchenko, {\it The Euler beta-function, the Vandermonde
determinant, the Legendre equation, and critical values of linear
function
on a configuration of hyperplanes,} I, Math. USSR-Izv. {\bf 35} (1990),
543-571; II, Math. USSR-Izv. {\bf 36} (1991), 155-168.
\item{[V2]} A.Varchenko {\it Determinant formula for Selberg type
integrals,} Funct. Analysis and its
Appl. {\bf 25} (1991), no.4, 88-89.
\item{[V3]} A.Varchenko, {\it Critical values and determinant of
periods,} Uspehi Mat. Nauk {\bf 44} (1987), 235-236.
\item{[V4]} A.Varchenko,  {\it Multidimensional hypergeometric
functions and representation theory of Lie algebras and quantum groups},
Preprint, 1992.

\end